\documentclass[preprint, superscriptaddress,amsmath, nofootinbib]{revtex4-1}
\usepackage{graphicx}
\usepackage{latexsym}
\usepackage{slashed}
\usepackage{bm}
\usepackage[dvipsnames,x11names]{xcolor}
\usepackage[colorlinks,citecolor=blue,urlcolor=blue,linkcolor=red]{hyperref}
\usepackage{diagbox}
\usepackage{makecell}
\usepackage{hyperref}
\pdfoutput=1
\usepackage{dcolumn}
\usepackage{bm}
\usepackage{multirow}
\usepackage{subcaption}
\usepackage{ulem}
\usepackage{tabularx}
\usepackage{hyperref}
\usepackage{cleveref}
\usepackage{caption}
\usepackage{ragged2e}  
\usepackage{booktabs}
\usepackage{amsmath,amssymb}
\usepackage{geometry}
\usepackage{hyperref}
\usepackage{booktabs}
\usepackage{microtype}
\definecolor{purple}{RGB}{128, 0, 128}

\begin{document}

\title{Searching for Dark Photon Tridents Through Primordial Black Hole Signatures}
\def\slash#1{#1\!\!\!/}

\author{Kingman Cheung}
\email{cheung@phys.nthu.edu.tw}
\affiliation{Department of Physics, National Tsing Hua University, Hsinchu 30013, Taiwan} 
\affiliation{Center for Theory and Computation,
National Tsing Hua University, Hsinchu 30013, Taiwan}
\affiliation{Division of Quantum Phases and Devices, School of Physics, Konkuk University, Seoul 143-701, Republic of Korea}

\author{C.J. Ouseph}
\email{ouseph444@gmail.com}
\affiliation{Institute of Convergence Fundamental Studies, Seoul National University
of Science and Technology, Seoul 01811, Korea}

\author{Po-Yan Tseng}
\email{pytseng@phys.nthu.edu.tw}
\affiliation{Department of Physics, National Tsing Hua University, Hsinchu 30013, Taiwan} 
\affiliation{Center for Theory and Computation,
National Tsing Hua University, Hsinchu 30013, Taiwan}
\affiliation{Physics Division, National Center for Theoretical Sciences,
Taipei 106319, Taiwan}

\author{Sin Kyu Kang}
\email{skkang@seoultech.ac.kr}
\affiliation{Institute of Convergence Fundamental Studies, Seoul National University
of Science and Technology, Seoul 01811, Korea} 
\affiliation{School of Natural Science, Seoul National University
of Science and Technology, Seoul 01811, Korea}

\date{\today}

\begin{abstract}

The detection of gamma-ray signals from primordial black holes (PBHs) could provide compelling evidence for 
their role as a dark matter candidate, particularly through the observation of their Hawking radiation. 
Future gamma-ray observatories, such as e-ASTROGAM, and the next-generation telescopes, are 
poised to explore this possibility by measuring both Standard Model (SM) and beyond-the-SM particle 
emissions. A particularly promising avenue involves production of dark photons by PBHs, which is a 
hypothetical particle that decays into photons.  
In this work, we investigate the trident decay of dark photons with mass $m_{A'}\leq 1$ MeV
focusing on their primary emission from asteroid-mass PBHs. We assume that the dark photons produced via Hawking radiation 
decay into photons well before reaching Earth, thereby enhancing the detectable gamma-ray flux. 
The energy spectrum of the photons decaying from the dark photons is distinct 
from that of direct Hawking-radiated photons due to higher degree of freedom, 
leading to observable modifications in the gamma-ray signal. 
Using the asteroid-mass PBHs as a case study, we demonstrate that future gamma-ray missions 
could detect dark-photon signatures and distinguish them from conventional Hawking radiation. 
This approach enables the exploration of previously inaccessible parameter spaces in dark photon mass
$m_{A'}\leq 1$ MeV and their coupling to photons, 
offering a viable avenue
to uncover 
the properties of dark sectors and the nature of asteroid-mass PBHs.

\end{abstract}

\maketitle
\section{Introduction}\label{Sec.1}

PBHs are considered  one of the macroscopic dark matter (DM) candidates, 
accounting for all or a fraction of the cold relic density~\cite{Hawking:1971ei,Chapline:1975ojl,Khlopov:2008qy,Carr:2016drx,Carr:2020gox,Carr:2020xqk,Green:2020jor}. They can be produced in the early universe through various scenarios:
(i) Collapse of overdense regions developed from primordial fluctuations after inflation~\cite{Carr:1974nx,Sasaki:2018dmp,Cheung:2023ihl},
(ii) invoking the first-order phase transitions, accumulating sufficient energy density 
within the Schwarzschild radius from bubble wall collisions~\cite{Hawking:1982ga,Kodama:1982sf,Moss:1994iq,Konoplich:1999qq}, and
(iii) incorporating dark sector particles in which PBHs are formed from the collapse of 
the macroscopic intermediate states called 
fermi-balls~\cite{Baker:2021nyl,Gross:2021qgx,Kawana:2021tde,Marfatia:2021hcp}.
Once the PBH mass falls below $10^{-15}$ solar mass, Hawking radiation becomes 
significant~\cite{Hawking:1975vcx,Gibbons:1977mu}, causing PBHs to emit both SM and beyond the Standard Model (BSM) particles. 
The emitted particle spectra depend only on the PBH temperature and the masses and spins of the particles, regardless of their interaction strengths. Therefore, PBHs are advocated as a candidate for producing feebly-interactive BSM particles, which is difficult to produce 
or detect in collider experiments. Many earlier works have explored PBH production of DM~\cite{Bell:1998jk,Allahverdi:2017sks,Lennon:2017tqq,Marfatia:2022jiz,Kim:2023ixo,Ray:2021mxu,
Gondolo:2020uqv,
Sandick:2021gew}, 
dark radiation~\cite{Arbey:2021ysg,Masina:2021zpu}, 
axion-like particle (ALP)~\cite{Schiavone:2021imu,Bernal:2021yyb,Mazde:2022sdx}, 
baryon asymmetry in the early 
universe~\cite{Carr:1976zz,Toussaint:1978br,Turner:1979bt,Baumann:2007yr,Fujita:2014hha,Hook:2014mla,Hooper:2020otu,Hamada:2016jnq}. PBHs can produce other feebly-interacting particles, e.g., sterile neutrinos~\cite{Chen:2023lnj,
Chen:2023tzd}, yielding X-ray + GW signals. Our dark photon trident is distinguishable via its unique $3\gamma$ spectrum, without charged lepton bremsstrahlung.

%

In this work, we focus on production of dark photons by Hawking radiation from PBHs.
The dark photon $A'$ is a hypothetical massive spin-1 particle,
which could originate from a $U(1)$ gauge extension of the SM. 
Through the kinetic mixing with the SM photon, the dark photon weakly couples to other SM particles. 
When the dark-photon mass exceeds twice the electron mass, $A'$ decays into 
a pair of $e^+ e^-$. If the mass is below the $e^+ e^-$ threshold, 
the dark photon decays into 3 photons. This occurs because
the tree-level di-photon decay channel is forbidden by the Landau-Yang 
theorem~\cite{PhysRev.77.242}, and so the loop-level photon trident channel 
$A' \to 3 \gamma$ becomes the dominant one~\cite{Linden:2024fby}.
The three-photon spectrum in this decay is distinguishable from conventional two-body final states. 
In contrast, this photon-trident signal might not be accessible in collider experiments
due to (i) all other existing constraints that force the kinematic mixing $\epsilon$ 
to be too small for practical production and detection, 
(ii) loop suppression in the trident decay, and 
(iii) difficulty in resolving 3 photons for dark photon mass below $2m_e$.
However, by considering the photon spectrum from the dark photons $A'$ produced 
by Hawking radiation from PBHs, we can overcome these challenges.

 Dark photons, as gauge bosons of an additional $U(1)_{\text{hidden}}$ symmetry, play a critical role in many BSM frameworks by serving as a bridge between visible and hidden sectors. Their interaction with the SM photon occurs through kinetic mixing, a renormalizable dimension-4 operator induced by loops of heavy particles charged under both gauge groups \citep{Holdom:1985ag}. This mechanism positions dark photons as a key portal for exploring hidden sectors, potentially mediating interactions in scenarios like self-interacting or inelastic dark matter, and facilitating processes such as freeze-in production. Their relevance is further underscored by their potential to resolve observational anomalies, such as the white dwarf cooling excess and the electron recoil excess observed at XENON1T \citep{Keung:2020uew}. With ongoing and planned experiments at facilities like NA64, LDMX, HPS, and FASER, Belle II \cite{Cheung:2022kjd,
Cheung:2024oxh,NA482:2015wmo,Araki:2020wkq,Chakrabarty:2019kdd,Graham:2021ggy} probing a broad mass range, dark photons remain a vital target for uncovering new physics.

PBH evaporation produces a distinctive gamma-ray signature via dark photon trident decays ($A' \to 3\gamma$), yielding a unique three-photon spectrum in the MeV range (0.1–10 MeV) that stands out against astrophysical backgrounds. Unlike the two-photon decays of spin-0 particles (e.g., $\pi^0 \to 2\gamma$), this spectrum exhibits a characteristic energy distribution, which cannot be replicated by smooth diffuse emission or power-law sources like pulsars or active galactic nuclei. The signal’s alignment with the Galactic dark matter halo’s spatial profile and its temporal stability further distinguish it from isotropic or transient backgrounds. For PBH masses of $10^{15}$ to $10^{17}$ g, the signal achieves a test statistic (TS) $> 9$ for instruments like e-ASTROGAM, corresponding to approximately $3\sigma$ significance, making PBH evaporation a robust and detectable probe of new physics in the MeV range.
In addition to e-ASTROGAM, Compton Spectrometer and Imager (COSI) experiment
\cite{Tomsick:2023aue} will be launched soon,
 which is the successor to COMPTEL.
 Although its projected
sensitivity is 1--2 orders of magnitude weaker than that of e-ASTROGAM and it
covers a narrower energy range, COSI may still provide a complementary near-term probe
of part of the PBH-induced dark-photon parameter space, as illustrated in Fig.~\ref{fig:1}.

We compute the photon spectrum from PBHs in the region within $|R| < 5^{\circ}$ of the Galactic Center
for two scenarios. 
We refer to (i) {\it SM-scenario}, where only SM particles are emitted via the Hawking radiation, and 
(ii) {\it dark-photon-scenario}, as the scenario where SM particles and dark photons are emitted via the Hawking radiation.
In the SM-scenario, the photon spectrum produced from PBH Hawking radiation includes 
primary photons, from neutral pion decays, electron final state radiation (FSR), and muon decay+FSR 
etc.~\cite{Arbey:2021mbl}. 
In the dark-photon-scenario, dark photons are produced in the primary emission from PBHs, and thus the photon-trident spectrum from dark photon decays contributes to the total photon spectrum. 
We find that the photon spectrum in the dark-photon-scenario can dramatically deviate from 
that of the SM-sceanrio when the dark-photon mass is lighter than the PBH temperature. 
Specifically, we adopt the PBH masses in the range $[10^{15}\,{\rm g},10^{18}\,{\rm g}]$ and 
use the sensitivity of e-ASTROGAM~\cite{e-ASTROGAM:2016bph,Agashe:2022jgk} to perform the Likelihood analysis.
As a result, we are able to identify the parameter space where the
dark-photon-scenario can be differentiated from the SM-sceanrio.

This work is organized as follows. In Section~\ref{Sec.2}, we review the dark photon with kinetic mixing to SM photon and formulate the trident-photon decay process. In Section~\ref{Sec.3}, we incorporate dark photon into the PBH Hawking radiation and derive the integrated gamma-ray spectrum from the Galactic Center. Section~\ref{Sec.4} utilizes the near future e-ASTROGAM sensitivity to perform a statistical analysis for the identification and discovery potential of the SM scenario and the dark-photon-scenario. Finally, we summarize in Section~\ref{Sec.5}. 
We include an appendix to estimate the extragalactic gamma rays,
which turns out negligible.

\section{Dark Photon Model}\label{Sec.2}

In this study, we expand the scope of gamma-ray searches for BSM physics, shifting the focus 
from PBHs alone to including new particles generated by PBHs, with 
the dark photon serving as a specific example. Specifically, we investigate the signal 
in the dark-photon-scenario, where dark photons are emitted from PBHs via Hawking radiation, 
followed by their decays into 3 photons. This process is added to the
direct photon emission from PBHs, resulting in a modification of the total spectrum.
We then contrast this signal with the SM-scenario in which PBHs produce only SM particles. 
Since observations of galactic gamma-ray signals impose stricter constraints 
on PBH abundance compared to studies of dwarf spheroidal galaxies~\cite{Coogan:2020tuf}, 
we concentrate on the Milky Way's gamma-ray signal for the dark-photon-scenario.

\noindent\textbf{\emph{Dark Photon Trident.}} --- The Lagrangian for the 
dark photon $A^{\prime}$ with mass $m_{A^{\prime}}$ and for kinemtic mixing 
with the SM photon is described by
\begin{equation}
    \mathcal{L} \supset -\frac{1}{4}F_{\mu\nu}^{\prime}F^{\prime\mu\nu}-\frac{1}{2}m_{A^{\prime}}^{2}A_{\mu}^{\prime}A^{\prime\mu}-\frac{\epsilon}{2}F^{\prime}_{\mu\nu}F^{\mu\nu},
    \label{eq:1}
\end{equation}
where the field strength tensor for the dark photon is defined as 
$F^{\prime}_{\mu\nu} = \partial_{\mu} A^{\prime}_{\nu} - \partial_{\nu} A^{\prime}_{\mu}$,
and $F^{\mu\nu}$ is the field strength for the SM photon.
The presence of a nonzero kinetic mixing parameter $\epsilon$ allows the dark photon to 
interact with other SM particles via mixing with the photon. The gauge symmetry breaking mass term for the dark photon can arise from a dark Higgs mechanism  \cite{Mondino:2020lsc,Carena:2004xs} or the Stueckelberg model (see e.g. Ref.~\cite{Cheung:2007ut}).

For the dark photon mass below twice the electron mass, the dominant decay mode is 
into three photons, a process termed the dark-photon trident. The only other 
kinematically feasible decay channel is into a neutrino pair, 
arising from dark-photon mixing with the SM $Z$ 
boson~\cite{Nguyen:2022zwb,Nguyen:2023ugx,Nguyen:2024kwy,Linden:2024fby,Linden:2024uph}. 
However, due to the large mass of the $Z$ boson, this decay is severely suppressed and 
thus neglected in this analysis.

The total decay width of a dark photon into three massless final-state particles can be expressed as ~\cite{McDermott:2017qcg}:
\begin{equation}
\begin{split}
\Gamma_{A^{\prime}\to\gamma\gamma\gamma} &= \frac{1}{(2\pi)^3} \frac{1}{32 m_{A^{\prime}}^3} \frac{1}{6} \\
&\times \int\limits_{0}^{m_{A^{\prime}}^{2}} dm_{12}^{2} \int\limits_{0}^{m_{A^{\prime}}^{2} - m_{12}^{2}} dm_{13}^{2} \, \overline{|\mathcal{M}|}^2,
\end{split}\label{eq:3.1}
\end{equation}
where $\mathcal{M}$ denotes the decay amplitude, which depends on the Dalitz variables $m_{ij}^2 = (k_i + k_j)^2$, constructed from the photon momenta $k_i$ and $k_j$ of the final state. The prefactor $1/6$ corrects for the presence of three identical bosons in the final state. 

The energy spectrum of photons produced in dark-photon trident decay is 
given by ~\cite{Pospelov:2007mp, Linden:2024uph}:
\begin{equation}\label{eq:5}
    \frac{{\rm d}N_{\gamma}}{{\rm d}E_{\gamma}} = \frac{2x^{3}}{17 m_{A^{\prime}}} (1715 - 3105x + \frac{2919}{2}x^{2}),
\end{equation}
where $x = 2E_{\gamma}/m_{A^{\prime}}$ varies between 0 and 1.

The kinematic distribution of the dark photon in the laboratory(lab) frame  
can obtained using the following formula~\cite{Berger:2019aox}. 
\begin{equation}\label{eq:6}
    \frac{dN'_\gamma}{dE'_\gamma} = \frac{1}{2} \int dE_\gamma \, d\cos\theta \, \frac{dN_\gamma}{dE_\gamma} \delta \left(E'_\gamma - \gamma(E_\gamma + \beta p \cos\theta)\right),
\end{equation}
\begin{equation}\label{eq:7}
    = \frac{1}{2} \int_{E'_1}^{E'_2} dE_\gamma \frac{dN_\gamma}{dE_\gamma} \frac{1}{p \beta \gamma},
\end{equation}
where \( E'_2 = \gamma (E'_\gamma + p'\beta) \) and \( E'_1 = \gamma (E'_\gamma - p'\beta) \) and the prime variables refer to those in the lab frame. This formula allows us to obtain the boosted spectrum from the decay spectrum at rest.

\section{Hawking Radiation Emission from Primordial Black Holes}\label{Sec.3}

\begin{figure}[h!]
    \begin{center}
    \begin{subfigure}{0.49\textwidth}
        \centering
        \includegraphics[width=\linewidth,height=6cm]{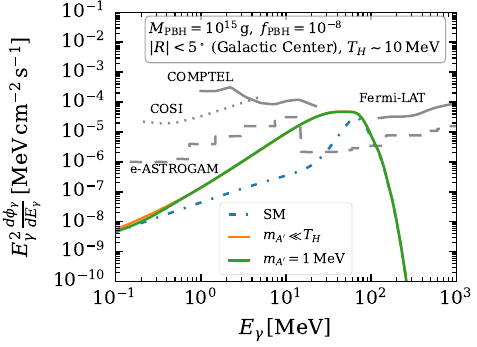}
    \end{subfigure}
    \hfill
    \begin{subfigure}{0.49\textwidth}
        \centering
        \includegraphics[width=\linewidth,height=6cm]{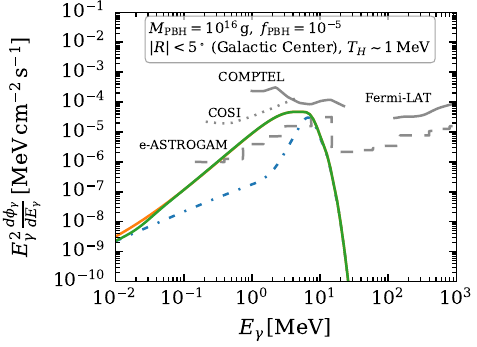}
    \end{subfigure}
    
    \vspace{1em} 
    
    \begin{subfigure}{0.49\textwidth} 
        \centering
        \includegraphics[width=\linewidth,height=6cm]{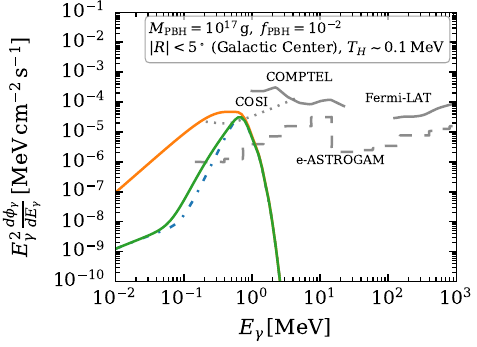}
    \end{subfigure}
\end{center}
    \caption{
 Gamma-ray spectra from Hawking radiation for PBHs with \( (M_{\rm PBH},f_{\rm PBH})=(10^{15} \, \text{g},10^{-8})\) in the top-left panel, \((10^{16} \, \text{g},10^{-5})\) in the top-right, and   \( (10^{17} \, \text{g},10^{-2})\) in the bottom-center. The spectra include the SM-scenario (dotted-dashed blue) and the dark-photon-scenario for different dark photon masses (various colors). The ROI is the Galactic center with \( |R| \leq 5^{\circ} \). Experimental constraints and future sensitivities are shown for COMPTEL (solid), Fermi-LAT~\cite{Fermi-LAT:2018pfs} (solid), COSI (dotted)~\cite{Tomsick:2023aue} and e-ASTROGAM~\cite{e-ASTROGAM:2016bph,Agashe:2022jgk} (dashed).}
\label{fig:1}
\end{figure}

A BH is predicted to continuously emit particles near its event horizon, a phenomenon known as Hawking radiation~\cite{Hawking:1974rv}. In this section, we discuss the particle spectra associated with Hawking radiation~\cite{Coogan:2020tuf}. Particles emitted directly from the BH are referred to as primary particles, whereas those resulting from interactions of primary particles are termed secondary particles. The production rate of a primary particle $i$ per unit time and energy from a BH of mass $M$ is given by ~\cite{Hawking:1974rv, Page:1976df, MacGibbon:1990zk}:
\begin{equation}\label{eq:8}
    \frac{\partial N_{i,\textrm{primary}}}{\partial E_i \partial t} = \frac{g_i}{2 \pi}\frac{\Gamma_i(E_i,M,m_i)}{e^{E_i/T_H}\pm 1},
\end{equation}
where $m_i$ and $g_i$ represent the mass and degrees of freedom of the particle $i$, 
$\Gamma_i$ is the graybody factor, and $T_H = 1/(8 \pi G M)$ is the Hawking temperature. The plus and minus signs correspond to fermions and bosons, respectively. The graybody factors describe the probability that a particle which generated at the horizon of a BH, propagate to spatial infinity. They are obtained by solving the particle's equation of motion in curved spacetime with specific boundary conditions.
At high energies, the graybody factor approaches the geometrical optics limit, $\Gamma_i = 27 G^2 M^2 E^2_i$. In this work, the exact graybody factor $\Gamma_i$  of non-rotating BH is computed using the \texttt{BlackHawk v2.0} package ~\cite{Arbey:2019mbc, Arbey:2021mbl}. \texttt{BlackHawk v2.0} includes the effect of the massive particle and imposes a cutoff on the evaporation spectrum for $E_i < m_i$ when the energy smaller than particle rest mass.

We focus on the photon spectrum of Hawking radiation, which includes contributions from 
both primary photons and secondary photons produced via decays and FSR of primary particles. The total photon spectrum is expressed as:
\begin{eqnarray}\label{eq:9}
\frac{\partial N_{\gamma,\textrm{tot}}}{\partial E_\gamma \partial t} &=& \frac{\partial N_{\gamma,\textrm{primary}}}{\partial E_\gamma \partial t} 
+ \int d E_{A^{\prime}}~\frac{\partial N_{{A^{\prime}},\textrm{primary}}}{\partial E_{A^{\prime}} \partial t} \frac{d N_{{A^{\prime}},\textrm{decay}}}{dE_\gamma} \nonumber \\
&& + \int d E_{\pi^{0}}~2\frac{\partial N_{\pi^{0},\textrm{primary}}}{\partial E_{\pi^{0}} \partial t} \frac{d N_{\pi^{0},\textrm{decay}}}{dE_\gamma} \nonumber \\
&& + \sum_{i=e^\pm,\mu^\pm,\pi^\pm} \int d E_i \frac{\partial N_{i,\textrm{primary}}}{\partial E_i \partial t} \frac{d N_{i,\textrm{FSR}}}{dE_\gamma}.
\end{eqnarray}
where:
\begin{eqnarray}\label{eq:10}
    \frac{d N_{\pi^0,\textrm{decay}}}{dE_\gamma} &=& \frac{\Theta(E_\gamma-E_{\pi^0}^-) \Theta(E_{\pi^0}^+-E_\gamma)}{E_{\pi^0}^+-E_{\pi^0}^-},\\\label{eq:11}
    E_{\pi^0}^\pm &=& \frac{1}{2} \left ( E_{\pi^0} \pm \sqrt{E_{\pi^0}^2 - m_{\pi^0}^2} \right ),
\end{eqnarray}
and
\begin{eqnarray}\label{eq:12}
    \frac{d N_{i,\textrm{FSR}}}{dE_\gamma} &=& \frac{\alpha}{\pi Q_i}P_{i\rightarrow i\gamma}(x) \left [\log \left (\frac{1-x}{\mu_i^2} \right ) -1 \right ],\\\label{eq:13}
    P_{i\rightarrow i\gamma}(x) &=& \begin{cases} \frac{2(1-x)}{x}, & i=\pi^\pm \\
    \frac{1+(1-x)^2}{x}, & i=\mu^\pm, e^\pm \end{cases}.
\end{eqnarray}

where $x = \frac{2E_\gamma}{Q_i}$, $\mu_i = \frac{m_i}{Q_i}$, and $Q_i = 2E_i$. Note that we ignore the three-body decay of $\mu^{\pm}$ and $\pi^{\pm}$. These processes can be safely neglected because their masses are much larger than the energy range of interest.

We note that PBH-emitted electrons and positrons can also produce secondary gamma-rays 
via inverse Compton scattering~(ICS) off ambient photon fields. However, this contribution 
is subdominant for our setup for two reasons. First, MeV-scale electrons lose energy 
and diffuse over length scales of several kiloparsecs before producing ICS photons, 
significantly larger than the angular scale of our ROI ($\sim 0.7$~kpc for $|R| < 5^\circ$). 
The resulting ICS emission is therefore spatially diffuse and largely falls outside our 
signal region. Second, the typical energy of photons produced by ICS is far below the MeV range 
relevant for our analysis. In the Thomson regime, the upscattered photon energy is 
approximately~\cite{RybickiLightman,BlumenthalGould}
\[
E_{\gamma}^{\rm IC}\simeq \frac{4}{3}\gamma_e^2 E_{\rm ph}.
\]
For \(E_e\sim 1\text{--}10\,{\rm MeV}\), one has 
\(\gamma_e\simeq E_e/m_e\sim 2\text{--}20\). Thus, scattering on CMB photons with 
\(E_{\rm CMB}\sim 6\times10^{-4}\,{\rm eV}\) gives
\[
E_{\gamma}^{\rm IC}\sim 10^{-3}\text{--}10^{-1}\,{\rm eV},
\]
which is far below the MeV energy range relevant for COMPTEL, e-ASTROGAM, and COSI.

On the other hand, the energy spectrum of photons produced in dark-photon trident decay 
is given in Eq.~[\ref{eq:5}-\ref{eq:7}].

The photon flux observed near Earth is given by:
\begin{equation}\label{eq:14}
    \frac{d \Phi_\gamma}{d E_\gamma} = \bar{J}_D \frac{\Delta \Omega}{4 \pi} \int dM \frac{f_\textrm{PBH}(M)}{M} \frac{\partial N_{\gamma,tot}}{\partial E_\gamma \partial t},
\end{equation}
where $\bar{J}_D$ is the J-factor for decay, defined as:
\begin{equation}\label{eq:15}
    \bar{J}_D = \frac{1}{\Delta \Omega}\int_{\Delta \Omega} d\Omega \int_\textrm{LOS} dl \rho_\textrm{DM}.
\end{equation}
The dark matter distribution in the Milky Way halo is modeled using a Navarro--Frenk--White (NFW) 
profile~\cite{Navarro:1996gj}:
\footnote {The NFW is a cuspy profile $\rho \sim r^{-1}$ near the 
 Galactic Center while the Moore \cite{Moore:1999nt} 
 is a more cuspy profile with 
 $\rho \sim r^{-1.5}$. As shown in Eq.(14) of our paper, 
 the more cuspy at the center implies larger flux 
 would be coming from the center. 
 On the other hand, the more cored profile, 
 such as isothermal ($\rho \sim r^0$),
 at the center implies less flux coming from the center. 
 Nevertheless, the integrated flux for a certain ROI 
 also depends on the profile beyond the core region. We calculate the ratios of J-factor for these profiles, obtaining $\bar{J}_D|_{\rm Iso}/\bar{J}_D|_{\rm NFW}=0.427$ and $\bar{J}_D|_{\rm Moore}/\bar{J}_D|_{\rm NFW}=1.38$.
Since \(\Phi_\gamma \propto \bar J_D f_{\rm PBH}\), the projected \(f_{\rm PBH}\) reach would be weakened (strengthened) by a factor of about \(2.3\) (\(0.72\)) for the isothermal (Moore) profile. This level of variation does not qualitatively change our main conclusions.}

\begin{equation}\label{eq:16}
\rho_{\rm DM}(r)=\frac{\rho_s}{\frac{r}{r_s} \, (1+\frac{r}{r_s})^2} \Theta(r_{200}-r),
\end{equation}
with parameters $r_s = 11~{\rm kpc}$, $\rho_s = 0.838~{\rm GeV}/{\rm cm}^3$, $r_{200} = 193~{\rm kpc}$, and $r_\odot = 8.122~{\rm kpc}$\cite{deSalas:2019pee}. For a region of interest (ROI) within $|R| < 5^{\circ}$ of the Galactic Center, the J-factor is $\bar{J}_D = 1.597 \times 10^{26}~{\rm MeV} {\rm cm}^{-2} {\rm sr}^{-1}$, and the angular size is $\Delta \Omega = 2.39 \times 10^{-2} {\rm sr}$.
The $f_{\rm PBH}$ is defined by the fraction of PBH and dark matter energy densities at present time~\cite{Carr:2020gox}.

In this study, we consider a monochromatic PBH mass distribution, which can arise, 
for instance, from the collapse of Q-balls ~\cite{Flores:2021jas} or a first-order phase
transition~\cite{Jung:2021mku}. Assuming $f_{\rm PBH}(M) = f_{\rm PBH} \delta(M - M_{\rm PBH})$ 
Eq.~(\ref{eq:14}) simplifies to
\begin{equation}\label{eq:17}
    \frac{d \Phi_\gamma}{d E_\gamma} = \bar{J}_D \frac{\Delta \Omega}{4 \pi}  \frac{f_\textrm{PBH}}{M_{\rm PBH}} \frac{\partial N_{\gamma,\textrm{tot}}}{\partial E_\gamma \partial t}.
\end{equation}

In the dark-photon-scenario, the gamma-ray spectrum arises from two contributions: 
(i) direct photon emission from PBHs and (ii) secondary emission resulting from dark-photon decays. 
Figure~\ref{fig:1} illustrates the photon spectrum for this scenario, 
considering PBH masses of \( M_{\rm PBH} = 10^{15} \), \( 10^{16} \), and \( 10^{17} \, \text{g} \), 
PBH abundance fractions of \( f_{\rm PBH} = 10^{-8} \), \( 10^{-5} \), and \( 10^{-2} \), 
and a range of dark-photon masses. 

The energy of photons produced from dark-photon decays (\( E_\gamma \)) is typically lower than the
initial dark-photon energy (\( E_{A^{\prime}} \)). Furthermore, the production of dark photons with 
masses significantly exceeding the primary photon peak is suppressed. 
As a result, the visible dark-photon peak always appears to the left of the primary photon peak. 
The dark-photon spectrum exhibits a distinct peak location and spectral shape compared to the photon spectrum in the SM-scenario.

\subsection{Dark Photon Parameter Space}\label{Sec.3.1}

\begin{figure}
    \begin{center}
\includegraphics[width=0.8\linewidth,height=0.6\linewidth]{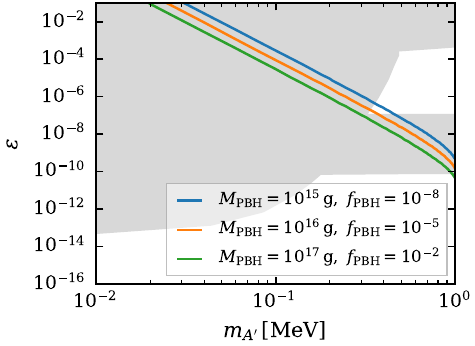}
    \end{center}
    \caption{The parameter space for dark photons that can be explored using PBHs is presented. The chosen PBH parameters are: \( M_{\rm PBH} = 10^{15} \, \text{g} \), \( f_{\rm PBH} = 10^{-8} \) (blue); \( M_{\rm PBH} = 10^{16} \, \text{g} \), \( f_{\rm PBH} = 10^{-5} \) (orange); and \( M_{\rm PBH} = 10^{17} \, \text{g} \), \( f_{\rm PBH} = 10^{-2} \) (green). Regions above the color curves are viable and existing constraints (gray) are taken from Ref.~\cite{Harnik:2012ni}}\textit{\textbf{}}. 
    \label{fig:2}
\end{figure}

For dark photons to alter the photon spectrum near the Earth, they must decay before arriving. 
The probability of dark photons decaying during their propagation from the Galactic center to Earth, assuming a monochromatic PBH mass, is given by~\cite{Agashe:2022phd}
\begin{equation}\label{eq:18}
    \langle P_{A^{\prime}, \textrm{decay}} \rangle\equiv \frac{\Phi_{A^{\prime},\textrm{dec}}}{\Phi_{A^{\prime},\textrm{tot}}},
\end{equation}
where
\begin{equation}\label{eq:19}
\Phi_{A^{\prime},\textrm{tot}} = \int_{\Delta\Omega} \frac{d\Omega}{4\pi} \int_{\rm LOS} d\ell \int dE_{A^{\prime}} \frac{f_{\rm PBH} \rho_{\rm DM}}{M_{\rm PBH}} \, \frac{\partial N_{A^{\prime},\textrm{primary}}}{\partial E_{A^{\prime}} \partial t} 
= \bar{J}_D \frac{\Delta \Omega}{4 \pi}  \frac{f_{\rm PBH}}{M_{\rm PBH}} \int dE_{A^{\prime}} \frac{\partial N_{A^{\prime}, \textrm{primary}}}{\partial E_{A^{\prime}} \partial t}
\end{equation}
\begin{equation}\label{eq:20}
\Phi_{A^{\prime},\textrm{dec}} = \int_{\Delta\Omega} \frac{d\Omega}{4\pi} \int_{\rm LOS} d\ell \int dE_{A^{\prime}} \frac{f_{\rm PBH} \rho_{\rm DM}}{M_{\rm PBH}} \, \frac{\partial N_{A^{\prime},\textrm{primary}}}{\partial E_{A^{\prime}} \partial t} \, P_{A^{\prime}, \textrm{decay}}(E_{A^{\prime}},\ell).
\end{equation}
The decay probability $P_{A^{\prime}, \textrm{decay}}$ is given by ~\cite{Agashe:2022phd}
\begin{equation}\label{eq:21}
P_{A^{\prime}, \textrm{decay}}(E_{A^{\prime}}, D)=1-\exp\left(-D~\Gamma_{A^{\prime}\to 3\gamma}\frac{m_{A^{\prime}}}{\sqrt{E^2_{A^{\prime}}-m^2_{A^{\prime}}}}\right),
\end{equation}
where $E_{A^{\prime}}$ is a nearly thermal energy distribution and $D$ is the
distance traveling from the PBH. 


We require $\langle P_{A^{\prime}, \textrm{decay}} \rangle$ to be greater than 99\% to 
determine the 
regions above the color curves in Fig.~\ref{fig:2} as viable ($\text{For each } m_{A^{\prime}}, \text{ the minimum } \epsilon~\text{that satisfies } P_{\text{decay}} \geq 99\% \text{ is determined}.$).  
For most of the parameter regions shown in Fig.~\ref{fig:2}, we assume the dark-photon decay 
is prompt since $\epsilon$ is at least one order of magnitude larger than the lower boundaries, which correspond to the color curves.
Since the decay width of dark photons is proportional to the square of the coupling, it follows that 
$\Gamma_{A^{\prime}\to 3\gamma} \propto \epsilon^2$. For the evaluation of $\Phi_{A^{\prime},\textrm{dec}}$ 
we assume that most of the dark photons originate from the Galactic 
center\footnote{$D$ in Eq.~\ref{eq:21} is fixed as the distance between the galactic center and Earth, approximately 8.3 kiloparsecs, which corresponds to a light travel time of $8.53 \times 10^{11}$ seconds. This is calculated using the speed of light ($c = 3 \times 10^8 \, \text{m/s}$) and 1 parsec $\approx 3.086 \times 10^{16} \, \text{m}$.}.
It is worth mentioning that, comparing with existing constraints (gray area in Fig.~\ref{fig:2}), this PBH dark-photon-scenario can probe the unexplored parameter space $10^{-7}\leq \epsilon \leq 10^{-3}$ and $m_{A'}\geq 0.5~{\rm MeV}$, i.e. the white trapezoidal area in Fig.~\ref{fig:2}, for which the $A'$ decay length is too long for terrestrial experiments meanwhile the mean free path is too short to escape from SN1987A. As a result, the PBH dark-photon-scenario provides a complementary sensitivity for dark photon.  While SN1987A bounds (e.g.,~\cite{Chang:2016ntp}) appear to exclude parts of this parameter space, uncertainties in supernova modeling allow order-of-magnitude variations, keeping the region marginally open. Recent X-ray constraints~\cite{Linden:2024fby} are complementary but do not fully close the window for PBH production.

\subsection{Dark Photon Decay Length and Nonlocal Effects}\label{Sec.3.2}
To ensure that the dark photons decay before reaching Earth, we require 
\begin{equation}
\langle P_{A^{\prime}, \textrm{decay}} \rangle > 99\% 
\end{equation}

as detailed in Eqs.~(\ref{eq:18})--(\ref{eq:21}). For our parameter space ($\epsilon \gtrsim 10^{-6}$--$10^{-4}$ for $m_{A'} \sim 0.5$--1 MeV), the rest-frame decay length is $c\tau \approx 6\times10^{15}$ m $\approx 0.0002$ kpc for $\epsilon=10^{-7}$. In the lab frame, the decay length is boosted by the Lorentz factor $\gamma \approx E_{A'}/m_{A'} \approx T_H / m_{A'} \approx 100$--200 (with $T_H \sim 10$ MeV for $M_{\rm PBH}=10^{15}$ g), yielding $l_{\rm lab} \approx 0.02$--0.04 kpc. This is much shorter than the distance to the Galactic Center ($\sim 8$ kpc) or the ROI scale ($\sim 1$--2 kpc).
Thus, the decays are effectively prompt at the source, and nonlocal effects on the emission morphology are negligible. This is supported by studies of nonlocal decay effects in similar scenarios~\cite{Jodlowski:2021xye}, which show that for decay lengths $\ll d_{\rm ROI} \sim 1$--2 kpc, the flux and morphology are nearly identical to the prompt case (see, e.g., Fig.~10 in Ref.~\cite{Jodlowski:2021xye}).
\section{Results \& Discussion}\label{Sec.4}

\begin{figure}
    \begin{center}
    \includegraphics[width=0.49\linewidth, height=7.5cm]{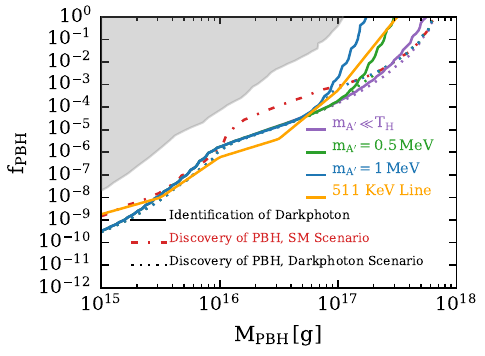}
    \hspace{0.000001\linewidth}
    \includegraphics[width=0.49\linewidth, height=7.5cm]{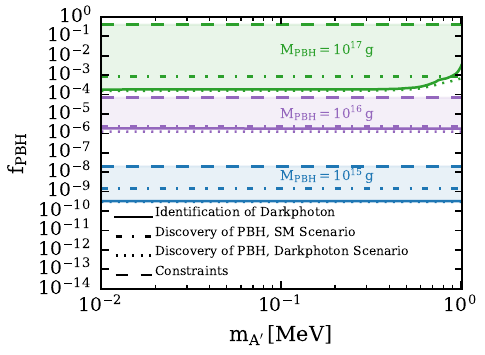}
    \end{center}
    \caption{
Bounds on PBH differentiability in the \( f_\mathrm{PBH} \) {\it vs.} \( M_\mathrm{PBH} \) plane ({\it left}) and the \( f_\mathrm{PBH} \) {\it vs.} \( m_{A^{\prime}} \) plane ({\it right}). 
The curves represent the ability to differentiate between a PBH signal in the SM-scenario ({\it dotted-dash}) and the dark-photon-scenario ({\it dotted}) from the background. 
Additionally, the distinguishability between the dark-photon-scenario and the SM-scenario is 
shown ({\it solid}), along with past experimental constraints 
({\it left: shaded light gray} and {\it right: dashed}). 
In the \( m_{A^{\prime}} \) plane, the shaded region indicates the parameter space where the dark-photon-scenario signal can be distinguished from the SM-scenario signal, with the far-right edge 
representing the maximum allowed \( m_{A^{\prime}} \).
}
    \label{fig:3}
\end{figure}

We use the likelihood analysis to constrain a specific model, 
assuming a Reference model that generates an observable gamma-ray 
signal~\cite{Agashe:2022jgk,Agashe:2022phd,Ouseph:2025ygv}, which, in our case, corresponds to the astrophysical background given
by experiments.
A Test model with inputs from the new physics is then evaluated against 
the Reference model. In our study, we consider two test models for PBH (i) the SM-scenario
and (ii) the dark-photon-scenario.
The probability that the Test model reproduces the 
gamma-ray signal predicted by the  Reference model follows the Poisson statistics 
represented as  
\begin{equation}
\mathcal{L}= \exp{\left(\sum_i n_i \ln{\sigma_i}-\sigma_i-\ln{n_i!}\right)},  
\label{eq.22}
\end{equation}  
where $n_i$ denotes the observed photon counts from the  Reference model, 
incorporating any background contributions, and $\sigma_i$ represents 
the expected photon count from the Test model including background 
within the $i$-th energy bin. 
To assess the relative validity of different Test models, the TS is 
introduced, defined as  
\begin{equation}
    \mathrm{TS} = - 2 \ln{\left(\frac{\mathcal{L}}{\mathcal{L}_\mathrm{Ref}}\right)} = \Sigma^2,  
    \label{eq:23}
\end{equation}  
where $\Sigma$ corresponds to the observational significance ~\cite{Cowan:2010js, Rolke:2004mj, 
Bringmann:2012vr, Fermi-LAT:2015kyq}, and $\mathcal{L}_\mathrm{Ref}$ is the likelihood 
associated with the Reference model. 
This analysis assumes that the combined statistical evaluation across gamma-ray energy bins follows 
a $\chi^2$ distribution. Unless stated otherwise, a significance threshold of 
$\Sigma = 3$ is adopted. 
The e-ASTROGAM foregrounds included hadronic, IC, and unresolved pulsar emissions. Our signal's unique spectrum, with DM-like morphology and stability, enables distinction; while future template fitting could further enhance this.

The construction of both the Reference and Test signals, along with the estimated background, 
is based on gamma-ray emission from the region around the Galactic Center, modeled 
using the NFW density profile within a $5^\circ$ field of view. 
The expected detector sensitivity and instrumental response are incorporated based 
on the design specifications of next-generation gamma-ray observatories. 
Additionally, forecasts of the astrophysical background are included to ensure 
a realistic evaluation of the observational prospects\cite{e-ASTROGAM:2016bph, Agashe:2022jgk}.\footnote{Please refer to Ref.~\cite{Agashe:2022jgk,Agashe:2022phd,e-ASTROGAM:2016bph} 
for more details about the detector sensitivity and foreground used.} 


\par In this work, we adopt a fixed astrophysical background model 
following the e-ASTROGAM instrument paper~\cite{e-ASTROGAM:2016bph}, whose sensitivity 
tables already incorporate a detailed model of the low-Earth orbit instrumental 
background; our results directly inherit those assumptions. Regarding astrophysical 
foreground uncertainties, we emphasize that the dark-photon scenario produces a 
spectral signature comprising a primary Hawking peak together with a displaced, 
softer dark-photon decay peak that is qualitatively distinct from all known smooth 
astrophysical foregrounds, including hadronic, inverse-Compton, and unresolved pulsar 
emissions. A spectral analysis exploiting this characteristic shape is therefore 
significantly more robust against foreground normalization uncertainties than a simple 
flux comparison. Furthermore, as visible from Fig.~\ref{fig:2}, the unexplored 
parameter space we identify spans several orders of magnitude in the kinetic mixing 
parameter $\epsilon$. We estimate that a factor-of-few uncertainty in the background normalization shifts 
sensitivity by at most $\mathcal{O}(1)$ in $\epsilon$, which does not qualitatively close 
the accessible window in the $(m_{A'}, \epsilon)$ plane. A full template-level marginalization over 
background model uncertainties using real e-ASTROGAM data remains an important avenue 
for future work.

\subsection{Identifying Signals and Differentiating Standard Model Processes from Dark Photon Interactions}\label{Sec.4.1}

To evaluate the detectability of a given model (PBH in the SM-scenario or PBH in dark-photon-scenario), 
we perform the aforementioned analysis, considering the 
astrophysical
background as the Reference model. 
Subsequently, we systematically scan the parameter space of the 
Test model to determine the values at which the 
likelihood deviates beyond a predefined significance level. The resulting constraints 
establish the PBH discovery limits, as depicted in Fig.~\ref{fig:3}. 
If the PBH population exceeds the corresponding threshold, the emitted signals 
will be sufficiently luminous to be detected above the background. Recent constraints from $e^+ e^-$ emission~\cite{DelaTorreLuque:2024qms} were based on cosmic-ray fluxes and the 511 keV line, complementary to our direct gamma-ray focus. Our benchmarks remain viable in the gamma-ray channel.

It is crucial to emphasize that for fixed \( M_\mathrm{PBH} \) and \( m_{A^{\prime}} \), 
the constraints in the dark-photon-scenario are at least as stringent as 
those in the SM-scenario. This is because PBHs produce the same SM particles, 
while the presence of dark photons introduces additional degrees of freedom 
(d.o.f =3 for dark photon), leading to an overall enhancement 
in signal brightness and detectability. 
Nevertheless, for heavier PBHs or heavier dark photons, the constraints in the dark-photon-scenario
asymptotically approach those of the SM-scenario. This transition occurs because the contribution 
from dark photons experiences exponential suppression when the Hawking temperature significantly 
falls below \( m_{A^{\prime}} \).

In the $( M_\mathrm{PBH},\, f_\mathrm{PBH} )$ plane presented in Fig.~\ref{fig:3}, 
the region above the solid curves delineates the parameter space where the dark-photon 
induced signal can be distinguished 
from the SM-scenario (i.e., the astrophysical background plus the SM-scenario as 
the Reference model).
Additionally, the figure illustrates the lower bounds on $f_\mathrm{PBH}$
required to differentiate a PBH signal from the background, with dotted lines 
representing the dark-photon-scenario and dot-dashed lines corresponding to the SM-scenario. As
expected, the threshold for distinguishing the dark-photon-scenario from the SM-scenario
is always higher than the detection threshold for the dark-photon signal itself, since the
presence of a detectable signal is a prerequisite for further characterization. Furthermore,
the identification threshold for the dark-photon signal remains consistently lower than that
of the SM-scenario due to its enhanced luminosity.

A notable feature is that for sufficiently large $M_\mathrm{PBH}$, the identification lines for the dark-
photon and SM scenarios converge. This convergence arises because dark photons become
exponentially suppressed when the Hawking temperature drops below their mass,
rendering their contribution negligible. Consequently, the distinguishability limits 
rapidly lose sensitivity, as the dark photon and SM signals become nearly indistinguishable. 
An additional factor contributing to this effect is the experimental sensitivity: 
at high PBH masses, the signal may extend beyond the detector's range, whereas in certain cases,
additional photons from dark-photon decays may populate regions of low detector sensitivity. 
For reference, existing PBH constraints in the SM case from Refs.~\cite{Clark:2018ghm, Green:2020jor, Coogan:2020tuf} are also included.

In the $( m_{A^{\prime}},\, f_\mathrm{PBH} ) $ plane of Fig.~\ref{fig:3}, the line 
styles remain consistent with those in the left panel. Solid curves denote the 
distinguishability limits between the dark photon and SM scenarios, whereas dotted lines 
represent the discovery limits for differentiating a PBH signal from the background 
in the presence of dark photons. For small \( m_{A^{\prime}} \), these curves remain relatively flat,
as dark photons behave effectively as massless particles due to their high energies. 
Conversely, for large \( m_{A^{\prime}} \), they transit to a nonrelativistic regime and 
undergo exponential suppression once the Hawking temperature falls below their mass. 
Consequently, the background identification threshold saturates, 
aligning with the SM-scenario identification curve (dot-dashed).

In contrast, the distinguishability curve between the dark photon and SM scenarios rises 
steeply at large \( m_{A^{\prime}} \) due to the diminishing contrast between the two signals.
Additionally, existing constraints on PBHs in the SM-scenario are depicted using 
dashed lines, indicating the region above which PBHs in the SM-scenario have 
already been excluded. The shaded regions highlight the viable parameter 
space where the dark-photon-scenario remains distinguishable from the SM-scenario.

\section{Conclusion}\label{Sec.5}
This study has investigated the detectability of gamma-ray signals from asteroid-mass PBHs, 
with a particular focus on the emission of dark photons with mass $m_{A'}\leq 1$ MeV via Hawking radiation. 
By analyzing the trident decay of dark photons, we establish that these events lead 
to distinct gamma-ray signatures, providing a means to differentiate the scenario 
from the conventional Hawking radiation. Our results indicate that future gamma-ray 
observatories, such as e-ASTROGAM, may be capable of probing previously 
unexplored regions of the dark-photon parameter space using asteroid-mass PBHs 
as a benchmark scenario.

The study has further demonstrated that the presence of dark photons enhances the gamma-ray flux, 
thereby improving the detectability of PBH signals. The influence of 
spin-dependent graybody factors significantly alters the energy spectrum of the emitted photons,
contributing to observable modifications in the gamma-ray signal. This effect enables 
a clear distinction between PBH emissions within the SM and those 
incorporating dark photons, reinforcing their viability as probes of the dark sector.

 We have also checked the possible extragalactic contribution to the gamma-ray signal. 
As shown in Appendix~A, within our Galactic-center ROI \((|R|<5^\circ)\), the 
extragalactic contribution is only about \(0.8\%\) of the Galactic-center flux. 
Therefore, it has a negligible impact on our results.

Furthermore, we show that the constraints imposed by dark photon emission are generally 
stronger than those from the SM scenario, except in the limit of large PBH masses 
or heavy dark photons, for which the signals converge due to exponential suppression. 
These findings offer a compelling case for future experimental efforts to search 
for dark photons in PBH emissions, presenting a novel pathway to explore new physics BSM.

Overall, this work provides a crucial step toward understanding the interplay between 
PBHs, dark photons, and the nature of dark matter. Future observational missions 
will be instrumental in testing these predictions, potentially uncovering 
signatures of new physics that extend beyond current theoretical frameworks.


\section*{Acknowledgment}
Special thanks to Yuhsin Tsai for the early discussions on gamma-ray spectrum-related calculations and to Nguyen Tran Quang Thong for valuable discussions.
K.C. is supported by the National Science \& Technology Council under grant no. NSTC 113-2112-M-007-041-MY3. C. J. O. and S. K. K are supported by the National Research Foundation of Korea under grant RS-2023-NR076731. P. Y. Tseng is supported in part by the National Science and Technology Council with Grant No. NSTC-111-2112-M-007-012-MY3, and Physics Division of the National Center for Theoretical Sciences of Taiwan with Grant NSTC 114-2124-M-002-003.

\begin{appendix}
\section{Estimating the Extragalactic Dark Matter Contribution:
          Comparison with the Galactic Center $J$-factor}

\subsection*{1.\; Galactic Center $J$-factor}
The photon flux from PBH Hawking radiation scales as
\begin{equation}
  \frac{d\Phi_\gamma}{dE_\gamma}
  \;\propto\;
  \bar{J}_D \cdot \Delta\Omega ,
\end{equation}
where the $J$-factor for a decaying dark matter signal is defined
as~\cite{Evans:2016xwx,Geringer-Sameth:2014yza}
\begin{equation}\label{eq:JD_def}
  \bar{J}_D
  \;=\;
  \frac{1}{\Delta\Omega}
  \int_{\Delta\Omega} d\Omega
  \int_{\rm LOS} dl\;\rho_{\rm DM}(\mathbf{r}) .
\end{equation}
Using the Navarro--Frenk--White~(NFW) profile~\cite{Navarro:1996gj} with
parameters $r_s = 11\;\text{kpc}$, $\rho_s = 0.838\;\text{GeV\,cm}^{-3}$,
$r_{200} = 193\;\text{kpc}$, and $r_\odot = 8.122\;\text{kpc}$~\cite{deSalas:2019pee},
and integrating over the region of interest $|R|<5^\circ$, the paper gives
\begin{equation}\label{eq:JGC}
  \bar{J}_D^{\rm GC}
  \;=\;
  1.597\times10^{26}\;\text{MeV\,cm}^{-2}\,\text{sr}^{-1} ,
  \qquad
  \Delta\Omega
  \;=\;
  2.39\times10^{-2}\;\text{sr} .
\end{equation}
The effective Galactic flux factor is therefore
\begin{equation}\label{eq:GC_flux}
  \Phi^{\rm GC}
  \;\propto\;
  \bar{J}_D^{\rm GC}\cdot\Delta\Omega
  \;=\;
  1.597\times10^{26}
  \;\times\;
  2.39\times10^{-2}
  \;\approx\;
  3.82\times10^{24}
  \;\text{MeV\,cm}^{-2} .
\end{equation}

\subsection*{2.\; Extragalactic $J$-factor}
\subsubsection*{2.1\; Formula}
For a decaying dark matter signal the extragalactic (EG) $J$-factor
integrates the mean cosmological DM density along the past light-cone,
accounting for Hubble expansion and cosmological redshift of the photon
energy~\cite{Ullio:2002pj,Ibarra:2009dr,Ibarra:2013cra}:
\begin{equation}\label{eq:JEG_def}
  J_D^{\rm EG}
  \;=\;
  \frac{\bar{\rho}_{{\rm DM},0}}{4\pi}
  \int_0^{z_{\rm max}}
  \frac{c\,dz}{H(z)\,(1+z)} ,
\end{equation}
where $\bar{\rho}_{{\rm DM},0} = \Omega_{\rm DM}\rho_{{\rm crit},0}$ is the
mean DM density today, $H(z) = H_0\sqrt{\Omega_m(1+z)^3+\Omega_\Lambda}$ is
the Hubble parameter in a flat $\Lambda$CDM universe, and the factor
$(1+z)^{-1}$ accounts for the cosmological redshifting of the photon energy.
Note that, in contrast to the annihilation case where the flux scales as
$\rho_{\rm DM}^2$~\cite{Ullio:2002pj}, the decay flux scales only linearly
with $\bar{\rho}_{\rm DM}$, so Eq.~\eqref{eq:JEG_def} is independent of the
dark matter clustering and is free of halo-model
uncertainties~\cite{Ibarra:2013cra}.

\subsubsection*{2.2\; Cosmological parameters}
We adopt the Planck~2018 flat $\Lambda$CDM
values~\cite{Planck:2018vyg}:
\begin{align}
  H_0 &= 67.4\;\text{km\,s}^{-1}\,\text{Mpc}^{-1}
                   \;=\; 2.184\times10^{-18}\;\text{s}^{-1} , \\
  \Omega_m &= 0.315 , \qquad
  \Omega_\Lambda \;=\; 0.685 , \qquad
  \Omega_{\rm DM} \;\approx\; 0.265 .
\end{align}
The critical density today and the mean cosmological DM density are
\begin{equation}
  \rho_{{\rm crit},0}
  \;=\; \frac{3H_0^2}{8\pi G}
  \;\approx\; 8.53\times10^{-30}\;\text{g\,cm}^{-3}
  \;\approx\; 4.79\times10^{-6}\;\text{GeV\,cm}^{-3} ,
\end{equation}
\begin{equation}\label{eq:rho0}
  \bar{\rho}_{{\rm DM},0}
  \;=\; \Omega_{\rm DM}\,\rho_{{\rm crit},0}
  \;\approx\; 0.265 \times 4.79\times10^{-6}\;\text{GeV\,cm}^{-3}
  \;\approx\; 1.26\times10^{-6}\;\text{GeV\,cm}^{-3} .
\end{equation}

\subsubsection*{2.3\; Hubble length}
\begin{equation}\label{eq:hubble_length}
  \frac{c}{H_0}
  \;=\;
  \frac{3\times10^{10}\;\text{cm\,s}^{-1}}{2.184\times10^{-18}\;\text{s}^{-1}}
  \;\approx\; 1.37\times10^{28}\;\text{cm}
  \;\approx\; 4.45\;\text{Gpc} .
\end{equation}

\subsubsection*{2.4\; Dimensionless redshift integral}
Define the dimensionless integral
\begin{equation}\label{eq:Iz}
  \mathcal{I}(z_{\rm max})
  \;\equiv\;
  \int_0^{z_{\rm max}}
  \frac{dz}{\sqrt{\Omega_m(1+z)^3+\Omega_\Lambda}\;(1+z)} .
\end{equation}
We take $z_{\rm max} = 10$ as the effective upper limit; contributions from
higher redshifts are negligible due to the opacity of the universe to MeV
photons~\cite{Ibarra:2009dr}. Representative values of the integrand
$f(z) = \bigl[\sqrt{\Omega_m(1+z)^3+\Omega_\Lambda}\,(1+z)\bigr]^{-1}$
are listed in Table~\ref{tab:integrand}, and numerical integration gives
\begin{equation}\label{eq:Iz_result}
  \mathcal{I}(10) \;\approx\; 0.9184 .
\end{equation}

\begin{table}[h]
\centering
\caption{Values of the integrand
$f(z) = \bigl[\sqrt{\Omega_m(1+z)^3+\Omega_\Lambda}\,(1+z)\bigr]^{-1}$
at selected redshifts, using Planck~2018 parameters ($\Omega_m=0.315$, $\Omega_\Lambda=0.685$).}
\label{tab:integrand}
\begin{tabular}{ccc}
\hline
$z$ & $\sqrt{\Omega_m(1+z)^3+\Omega_\Lambda}$ & $f(z)$ \\
\hline
0.0  & 1.000  & 1.0000 \\
0.5  & 1.322  & 0.504  \\
1.0  & 1.790  & 0.279  \\
2.0  & 3.032  & 0.110  \\
5.0  & 8.290  & 0.0201 \\
10.0 & 20.49  & 0.00444 \\
\hline
\end{tabular}
\end{table}

\subsubsection*{2.5\; Numerical result}
Substituting Eqs.~\eqref{eq:rho0}--\eqref{eq:Iz_result} into
Eq.~\eqref{eq:JEG_def}:
\begin{align}
  J_D^{\rm EG}
  &= \frac{\bar{\rho}_{{\rm DM},0}}{4\pi}
     \cdot \frac{c}{H_0}
     \cdot \mathcal{I}(z_{\rm max})
  \nonumber\\[4pt]
  &= \frac{1.26\times10^{-6}\;\text{GeV\,cm}^{-3}}{4\pi}
     \;\times\; 1.37\times10^{28}\;\text{cm}
     \;\times\; 0.9184
  \nonumber\\[4pt]
  &\approx 1.26\times10^{21}\;\text{GeV\,cm}^{-2}
  \;=\; 1.26\times10^{24}\;\text{MeV\,cm}^{-2} .
  \label{eq:JEG_result}
\end{align}
Since the EG signal is isotropic, the flux arriving within the solid
angle $\Delta\Omega = 2.39\times10^{-2}\;\text{sr}$ of the ROI is
\begin{equation}\label{eq:EG_flux}
  \Phi^{\rm EG}
  \;\propto\;
  J_D^{\rm EG}\cdot\Delta\Omega
  \;=\;
  1.26\times10^{24}
  \;\times\;
  2.39\times10^{-2}
  \;\approx\;
  3.01\times10^{22}
  \;\text{MeV\,cm}^{-2} .
\end{equation}

\subsection*{3.\; Comparison: EG vs.\ GC}
\subsubsection*{3.1\; Flux ratio}
From Eqs.~\eqref{eq:GC_flux} and~\eqref{eq:EG_flux}:
\begin{equation}\label{eq:ratio}
  \frac{\Phi^{\rm EG}}{\Phi^{\rm GC}}
  \;=\;
  \frac{J_D^{\rm EG}}{\bar{J}_D^{\rm GC}}
  \;=\;
  \frac{1.26\times10^{24}}{1.597\times10^{26}}
  \;\approx\; 0.00789 .
\end{equation}
The EG flux is roughly 0.79\% of the GC contribution (about two orders of magnitude smaller) within the $5^\circ$ ROI. 
Thus, EG gamma-ray flux would not affect our results.

\end{appendix}

\bibliography{paper}
\end{document}